\documentclass[12pt]{article}
\usepackage[body={17.5cm, 22cm},right=2cm]{geometry}
\usepackage{amssymb,graphicx}
\usepackage{amsmath}
\usepackage{epsfig}
\usepackage{hyperref}

\def\eps{\epsilon}
\def\d{\partial}
\def\l{\left(}
\def\r{\right)}

\newcommand{\be}{\begin{equation}}
\newcommand{\ee}{\end{equation}}
\newcommand{\bea}{\begin{eqnarray}}
\newcommand{\eea}{\end{eqnarray}}
\newcommand{\bg}{\begin{gather}}
\newcommand{\eg}{\end{gather}}
\newcommand{\bseq}{\begin{subequations}}
\newcommand{\eseq}{\end{subequations}}

\def\half{\frac{1}{2}}

\newcommand{\ket}[1]{| #1 \rangle}
\newcommand{\p}{\partial_+}
\newcommand{\m}{\partial_-}
\begin{document}
\begin{center}
{\Large\bf Superluminal Travel, UV/IR Mixing and  Turbulence \\ \vspace{0.3cm}in the Lineland}\\
\vspace{0.5cm}
{ \large
 Sergei~Dubovsky, Victor Gorbenko}\\
\vspace{.2cm}
{\small  \textit{Center for Cosmology and Particle Physics, Department of Physics, \\New York University, New York, NY, 10003, USA }}

\end{center}
\begin{center}
\begin{abstract}
We study  renormalizable Lorentz invariant stable quantum field theories in two space-time dimensions with instantaneous causal structure (causal ordering induced by the
light ``cone" time ordering).  These models provide a candidate UV-completion of the two-dimensional ghost condensate.
They  exhibit a peculiar UV/IR mixing---energies of all excitations become arbitrarily small at high spatial momenta. We discuss several phenomena associated with this mixing.
These include the impossibility to reach a thermal equilibrium and metastability of all excitations towards decay into short wavelength modes resulting in an indefinite turbulent  cascade.
 In spite of the UV/IR mixing in many cases  the UV physics can still be decoupled from low energy phenomena. However, a patient observer
in the Lineland is able to produce arbitrarily heavy particles simply by waiting for a long enough time. 
\end{abstract}
\end{center}

\section{Introduction}
Our chances for superluminal travel are very slim.
There is a single Lorentz invariant causal structure in $(3+1)$ dimensional Minkowski space-time and it is not compatible with superluminal signal propagation.
However,  life may be more interesting in a $(1+1)$ dimensional world---the Lineland \cite{superlum}.
There one may use a light ``cone" coordinate 
\[
x^+=t+x
\]
as an absolute time.
Then one may define an unconventional causal structure, such that the causal future of any observer at the moment of time $x_+=a$ consists of all space-time events in the upper right half-plane
$x^+>a$, and the causal past consists of all space-time events in the lower left half-plane $x^+<a$. Lorentz boosts in two dimensions act as 
\[
x^+ \to \lambda x^+,\;x^- \to \lambda^{-1}x^-
\]
with $\lambda>0$, so that the light cone causal ordering is Lorentz invariant. Its existence is related to the absence of continuous spatial rotations in the Lineland. All what is left is a discrete spatial parity
$x^+\leftrightarrow x^-$, which is necessarily  broken with this causal ordering (as it is in the real world).
Following \cite{superlum} we call this causal structure instantaneous, since it does not admit any causally disconnected
regions. From now on we will be referring to $x^+$ and $x^-$ as time and space coordinates correspondingly, unless specified otherwise. 

As shown by  \cite{superlum}, the Einstein-aether theory in two dimensions  \cite{Jacobson:2000xp,Eling:2006xg} provides an example of a renormalizable (in fact, asymptotically free) and apparently consistent 
theory realizing this causal structure\footnote{To be precise, we are talking here only about vector sector of the Einstein-aether, with gravity being decoupled.}. This model is subject to the following subtlety though. At the classical level the Einstein-aether theory exhibits a spontaneous breaking of Lorentz symmetry. One may expect this symmetry to be restored at the quantum level as prescribed by the Coleman--Mermin--Wagner theorem \cite{Mermin:1966fe,Coleman:1973ci}. However, non-compactness of the Lorentz group complicates direct construction of 
well-defined Lorentz invariant observables in this case.

Instead, in the current paper we chose to concentrate on simpler examples of instantaneous theories such that this subtlety is avoided.
Our main goal is to develop a basic intuition about these models by performing a number of simple calculations, aimed to demonstrate that 
instantaneous theories are amenable to a straightforward perturbative analysis and possible to work with in spite of a number of rather
 unconventional and, at first sight, counterintuitive properties.
 
 The origin of these peculiarities is easy to understand. The Lorentz invariant dispersion relation
in instantaneous theories is
\be
\label{w=m/k}
\omega={\mu^2\over |k|}\;,
\ee 
where $\omega$ is a frequency defined with respect to the time coordinate which is $x^+$ and $k$ is a spatial momentum with respect
to $x^-$. This dispersion relation exhibits a peculiar form of the UV/IR mixing---the higher the spatial momentum of a mode is, the 
smaller is its energy! In particular, the spectrum is always gapless no matter how heavy the ``mass" $\mu^2$ of a field is.
Understanding the consequences of this unconventional dispersion relation will be our main theme.

%The behaviour of field theories obeing instantaneous causual structure is remarkably diffirent from that of usual ones and 
%the aim of this paper is to make the ordinaryl, very basic, steps to develope some intuition
%about the theories of this type using the example of massive self-interacting scalar field, 
%and we will not even get as far as loop corrections. 
  
The rest of the paper is organized as follows. In section~\ref{classical} we start by introducing 
the class of theories we will be working with. To keep things as simple as possible we restrict to models of a single scalar field $\phi$.
A Lorentz invariant kinetic term with two time derivatives is $(\p \m \phi)^2$. Note that 
it has four derivatives with respect to the  Minkowski time $t$, so a theory with this kinetic term 
would contain ghosts, if quantized with respect to the conventional causual structure.  This problem does not arise when quantizing using the light cone time and instantaneous causal structure.

With this kinetic term  the naive scaling dimension of the field  $\phi$ is $-1$, 
 so  there is an  infinite number of  relevant and marginal operators.  For example, any  interaction Lagrangian of the form 
$P(\phi,\p\phi \m\phi)$ leads to a renormalizable theory according to the naive power counting.
In general, this will lead to a strongly coupled theory in the IR ({\it i.e.}, at small $\omega$ and $k$), however, as we will see, this strong coupling can be avoided by introducing a sufficiently large ``mass" term $\mu^4\phi^2$. 
Our principal working model of an interacting instantaneous theory will be a $\phi^3$ theory, which is not only super-renormalizable, but also finite.

To study the physics of instantaneous theories we start by  looking at the non-interacting model.
We calculate the retarded propagator to see explicitly the presence of instantaneous signals and discuss an unusual consequence 
of the dispersion relation (\ref{w=m/k})---the impossibility to reach the equilibrium Bose--Einstein distribution at finite temperature.
We continue in section \ref{quantization} with introducing the cubic interaction and discuss another interesting consequence of the $\omega/k$ dispersion relation---the instability of all particles  towards a two body decay into particles with {\it larger} spatial momenta. We also argue that in spite of the UV/IR mixing  the approximate decoupling of heavy particles ({\it i.e.}, particles with large $\mu^2$) is still possible. Nevertheless, the
UV/IR mixing makes high energy experimental physics simpler and less expensive---an arbitrarily heavy particle will be eventually produced in the {\it decays} of lighter particles, so the discovery of the theory of everything
is simply a matter of time in the Lineland.

 In order to understand what the absence of the equilibrium Bose--Einstein distribution means, in
 section 
\ref{kinetical} we  
use the Boltzmann equation to 
describe the non-equilibrium behaviour of a homogeneous gas in the instantaneous $\phi^3$ theory.
As a consequence of particle decays an arbitrary initial state gives rise to a turbulent cascade towards higher momenta.
An arbitrary initial distribution function approaches the universal scaling shape at late times. In the limit of small occupation numbers 
we find this scaling solution analytically. Physically, this cascade can be interpreted as a process of Bose condensation which cannot be completed in a finite time, because zero energy state corresponds to the infinite spatial  momentum.
We present our conclusions in section
\ref{discussion}.

Before proceeding, a comment is in order. 
To ensure a better IR (or, to be more precise, long distance) behavior it is often convenient to compactify
the $x^-$-coordinate on a circle. One may complain that this breaks the Lorentz symmetry. However,
from our view-point, the major interest of working with a Lorentz invariant theory is that it can be coupled to gravity in a natural way (we leave
for a future work the study of what happens then). A ``soft" breaking of Lorentz symmetry, like imposing periodic boundary conditions in
$x^-$,  does not spoil this property.

%-----------------------------------------------------------------------------------------------------------------------------------------

\section{Free Theory}
\label{classical}
A simple example of a free instantaneous theory is provided by the following Lagrangian,
\be
{\cal L}= \half (\p \m \phi)^2  -\half \mu^4 \phi^2\;.
 \label{Lfree}
\ee
The corresponding field equation in the Fourier space is 
\be
 (\omega^2k^2 -\mu^4)\phi = 0
\ee
so that at any absolute value $|k|$ of the spatial momentum we find one left-moving and one right-moving mode with the
UV/IR mixing dispersion relation (\ref{w=m/k}). Massless instantaneous theories are somewhat degenerate already at the level of a free theory---the dispersion relation (\ref{w=m/k}) does not give rise to  propagating waves for $\mu=0$. In what follows we restrict to the massive case $\mu^4>0$.

In principle, one can also  include in (\ref{Lfree}) a conventional kinetic term $Z \mu^2 \d_+\phi\d_-\phi$. 
Then at small frequencies and momenta our model would reduce to a conventional scalar field theory quantized in the light front.
This opens an interesting possibility to interpolate smoothly between conventional and instantaneous physics.
For simplicity, we restrict to the case $Z=0$, this choice can be enforced by requiring the invariance under $x^-\to -x^-$ parity. 

A well-known subtlety in the light front quantization is the  zero mode problem (see, e.g., \cite{Hellerman:1997yu}). 
The cause of the problem is that if one expands the field $\phi$ into modes with different spatial momenta the constant $k=0$ mode does not acquire a kinetic term from (\ref{Lfree}).

It turns out this problem ameliorates in instantaneous theories due to the presence of the term in (\ref{Lfree}), which is second order in the light cone time. To see this it is convenient to compactify the light cone time on a finite interval and to impose zero Dirichlet boundary conditions.
This way the zero mode gets projected from the theory and does not lead to any additional constraints. One may recover the non-compact theory by
sending the compactification length to infinity.

Note, that this argument cannot be directly applied to conventional theories. 
First, compactification on the light-like direction is subtle with a conventional causal structure. Even if one ignores this, it is immediate to see that the Dirichlet boundary conditions are incompatible with the field equation in the absence of the second derivative kinetic term.
Indeed, given some initial data at any  moment of time one can think of the field equation as an ordinary differential equation in $x^-$ for the uknown 
function
 $
 f=\p^2 \phi$.
 This equation is second order for the theory (\ref{Lfree}), 
 \[
 \m^2 f = \mu^4 \phi\;,
 \]
 so that one can impose two Dirichlet boundary 
conditions. For the conventional theory this equation is first order so it is impossible to impose two Dirichlet conditions.

Clearly,  the same reasoning can be applied to project out the zero mode in the interacting theory as well.  

Note also that for a conventional theory without higher derivative terms the light cone and conventional quantizations give rise to the same results, if done properly. The theory (\ref{Lfree}) demonstrates this is not the case in general.
Treated by conventional methods this model contains negative energy states---ghosts. 
As a consequence, already at the classical level, if one adds non-linearity into this theory the vacuum becomes unstable, arbitrary small field fluctuations grow indefinitely by populating positive and negative energy modes. This does not happen if one uses the light cone time to define the Cauchy problem. 

One should not be surprised that different  choices of a time coordinate or,  equivalently, different ways to pose the Cauchy problem give rise to 
inequivalent physical results even though the Lagrangian is the same. A more familiar example of this kind is a conventional massive theory, which turns from being stable into tachyonic if one exchanges the role of space and time. With different choices of the Cauchy surfaces one imposes 
different regularity conditions at the infinity on the allowed field configuration, and this may tame/induce the instability of the Cauchy problem.
%We can think of the equation of motion at any given
%moment of time as of an ordinary differential equation for the unknown function
%$f=\p^2 \phi$, namely,
%$\m^2 f = -\mu^4 \phi$,
%where the rhs is considered as known. Since the equation is of the second order, two
%boundary conditions can be specified, naturally they are homogenious
%Dirichlet boundary conditions, which automatically exclude the zero mode, so
%no usual zero mode constraint arises here. Note that these considerations are not affected
%by any nonliniarities which can be added to the equation of motion.
%are present in the equation these c 
%The zero mode can be excluded from this theory (unless there are terms
%ike $\p\m \phi^2$) if we first consider it on a finite segment with , and then take the length of the segment to infinity. Having this in %mind,
%we will not discuss the well-known zero mode problem [...] here.

 To see explicitly the instantaneous behavior in the theory (\ref{Lfree}) let us study the field response to an external source, which is determined by the retarded propagator,
 \be
 \label{retarded}
  G_{ret}(x^+,x^-)=\frac{1}{4 \pi^2}\int dkd\omega \frac{e^{ikx_-+ i\omega x_+}}{k^2 \omega^2-\mu^4-i\eps \omega}=-\frac{\theta(x_+)}{2\mu^2}J_0(\sqrt{ \mu^2 x_+ |x_-|})
\ee
%The integral over $\omega$ gives 
%\be
%\label{integral}
% G_{ret}(x_+,x_-)= \theta(x^+) \frac{ i}{2\pi} \int dk \l \frac{e^{ikx_- + i\frac{\mu^2}{k}x_+}}{2k \mu^2}
% - \frac{e^{ikx_- - i\frac{\mu^2}{k}x_+}}{2k \mu^2}\r
%\ee
%This integral is an even function of $x^-$.  
%The second term in (\ref{integral}) is zero for positive $x_+,x_-$ and the first term is equal to 
% and that the whole integral 
%is even in $x_-$. We then proceed splitting the first term into two integrals over positive and negative $k$, and introducing 
%$\lambda=\sqrt{ \mu^2 x_+ x_-}$ and
%$\e^{\alpha}=\pm k\sqrt{x_-/x_+\mu^2}$:
%\be
% \frac{i}{2 \pi} \int d\alpha \frac{e^{i \lambda ch(\alpha)}-e^{-i\lambda ch(\alpha)}}{2 \mu^2}=-\frac{1}{2\mu^2}J_0(\lambda)
%\ee
%finally,
%\be
% G_{ret}=-\frac{\theta(x_+)}{2\mu^2}J_0(\sqrt{ \mu^2 x_+ |x_-|})
% \label{Gret}
%\ee
We see that there is indeed  a non-vanishing field response at all $x^+\geq0$. One may worry that the fall-off of the retarded propagator
at large $x^-$ becomes more and more slow at early times. In particular, at $x^+=0^+$ the retarded propagator approaches a non-vanishing constant, indicating that the $k=0$ infinite frequency mode gets excited.
This is an artefact of using the source which is  switched on and off instantaneously.  For physical sources smeared in time (and space) the field response goes to zero at the spatial infinity at all times.

Apart from instantaneous signal propagation the retarded propagator (\ref{retarded}) exhibits 
other interesting features following from the UV/IR mixing dispersion relation (\ref{w=m/k}) and persisting in the interacting theory.
Consider an observer located at a fixed spatial position $x^-$.
After a localized source turns on  at the origin, she immediately observes the field response.
At late times $x^+ \to \infty$ this response asymptotes to 
\[
G_{ret}(x^+\to \infty, x^-)\propto
(x^+ |x^-|)^{-1/4}\sin(\sqrt{ \mu^2 x^+ |x^-|})\;.
\]
We see that the characteristic wavelength, $\lambda\sim \mu^{-1}\l |x^-|/x^+\r^{1/2}$, of the observed signal becomes shorter at late time.
On the other hand the characteristic time-scale for its variation, $\tau \sim  \mu^{-1}\l x^+/|x^-|\r^{1/2}$, becomes very long. So at late times the observer detects a very short-scale and almost static ``noise".

Another piece of information which is usually straightforward to extract from a free theory is a finite temperature behavior of a system.
However, we find a surprise here.
A thermal theory has to possess the Bose-Einstein distribution
\be
  n(k)=\frac{ 1}{
  e^{
 \l\frac{\mu^2}{|k|}-\bar{\mu}\r /T}-1
  }\;,
 \label{BE}
\ee
where $\bar{\mu}<0$ is a chemical potenital.
However, the total particle density given by $\int n(k) dk $ diverges at large momenta.
 This is  another manifestation of the UV/IR mixing---there is an infinite number of high momentum levels 
 with energies arbitrary close to zero, so that one expects to observe a 
never ending process of Bose-Einstein condensation onto the infinite momentum ground state. In section~\ref{kinetical}
 we will provide a detailed kinetic description of this process.

Note, that 
fermions obeying the UV/IR mixing dispersion relation (\ref{w=m/k}) would exhibit the same property---changing the sign in the denominator of the distribution (\ref{BE}) does not make its integral converging. The physical reason for that is similar---for any finite Fermi
 momentum there is always an infinite number of states below the Fermi surface, so that thermalization cannot be completed.

%---------------------------------------------------------------------------------------------------------------------

\section{Interacting Theory}
\label{quantization}
Let us turn to an interacting theory now. For simplicity, we restrict to the simplest (super)re\-nor\-malizable interaction\footnote{As usual, the vacuum is unstable in the $\phi^3$-theory  at the non-perturbative level, but this is irrelevant for our perturbative analysis.} 
\[
V_{int}={g\over 6}\phi^3\;.
\]
It is immediate to see now the dynamics responsible for the absence of thermalization pointed out at the end of the previous Section.
Namely, a peculiar property of the dispersion relation (\ref{w=m/k}) is that two body decays of a particle into a pair of particles of the same mass
and {\it larger} momenta are kinematically allowed. 
For a particle with a momentum $k$ the momenta of the decay products are
%
%In fact, let us consider the process $\ket{k}\to \ket{p}\ket{q}$. The equations
%\be 
%  k=p+q, \mu^2/|k|=\mu^2/|p|+\mu^2/|q| 
%\ee 
\[
p= -{\varphi k},\;\; q=k-p={ \varphi^2 k}\;,
\]
 where $\varphi=  (\sqrt{5}+1)/2$ is the golden ratio. Consequently, the momenta of decay products in an infinite 
  chain of two body decays originated from a single initial particle can be approximated by the Fibonacci numbers.
 % To exhibit
%this property, we add the interaction term 
%$1/3! g\phi^3$
%to our theory and calculate the life time of the particles.
%The nonperturbative instability of this theory will not bother us in what follows, since we will restrict
%ourselves to the first order perturbation theory. Nevertheless, one can substitute the interaction term 
%$g ( \p \m \phi) \phi^2$ instead.
%The latter does not contribute to the energy, thus leaving the Hamiltonian positively-definite. 
%The first-order amplitudes, however, remain unchanged.

Let us calculate the corresponding decay rate. The calculation is parallel to what one does in conventional theories, 
one should only be careful to keep track of the correct state and operator normalizations.
We introduce canonical creation and anihilation operators $a_k^{\dagger},a_k$ satisfying
\[
[a_k,a^\dagger_{k'}]=\delta(k-k')\;.
\]
For $\phi$ and and its canonically conjugate momentum $\pi_\phi$ to satisfy the canonical commutation relations, the field operator should be defined as follows 
\be
  \phi(x)=\int \frac{dk}{(2 \pi)^{1/2}} \frac{e^{ikx_-+i\omega x_+}}{\mu \sqrt{2|k|}}a_k+c.c.
  \label{field}
\ee
Note the extra $|k|^{-1}$ factor in the wave function normalization compared to the standard $(2\omega)^{-1/2}$ \cite{Weinberg:1995mt}, it is
related to the presence of two spatial derivative in the expression for the canonical momentum, $\pi_\phi=-\d_+\d_-^2\phi$.
%The next step is to calculate the matrix element for the system enclosed in a space-time box $L \times T$. In the box:
%\be
%  \delta(k-k')  \to \frac{L}{2\pi} \delta_{kk''} , \quad   \delta(w-w')  \to  \frac{1}{2\pi}\int_{-T/2}^{T/2} e^{i(w-w')t} dt, \quad   
%  \int dk \to \sum_k \frac{2\pi}{L} 
%\ee
%and consequently one-particle state should be normalized as follows:
%\be
%  \ket{k}_{box}= \sqrt{\frac{1}{L}} a^{\dagger}_k \ket{0}
%\ee
%Then for the matrix element we get the following expression:
%\be
%  \bra {k }g\phi^3 \ket{pq}_{box}= \frac{g}{\mu^3} \frac{\delta_{k,p+q} L \int_{-T/2}^{T/2} e^{i(w(k)-w(p)-w(q))t} dt}
%                                                                                                 {\sqrt{|kpq|L^3}}
%\ee
%  Now the regularised decay rate is
%\be
%  \Gamma_{box} T = \frac{g^2}{\mu^6|k|} \sum_{p,q} \frac{1}{L^2} \frac{\delta_{k,p+q} L \int_{-T/2}^{T/2}       e^{i(w(k)-w(p)-w(q))t} dtT} {|pq|}
%\ee
%  and when the infinite box limit is taken we arrive at the result
From here the standard expressions for the Feynmann rules and transition rates \cite{Weinberg:1995mt} adopted for the theory at hand give the following result for the two-body decay width,
\be
\label{Gamma}
  \Gamma = \frac{g^2}{8\mu^6|k|}
                    \int dp dq \frac{\delta(k-p-q) \delta \left( \frac{\mu^2}{|k|}-\frac{\mu^2}{|p|}-\frac{\mu^2}{|q|} \right)}{|pq|}  
                = \frac{g^2}{8\mu^8|k|} \left( \varphi+\frac{1}{\varphi} \right)^{-1}
\ee
Note that the ratio of the width to the frequency of a particle $\Gamma/\omega$ does not depend on the spatial momentum $k$.
This behavior could have been anticipated from the Lorentz invariance of the theory, so that the result  (\ref{Gamma})
for the width $\Gamma$ is 
fixed by the Lorentz invariance and dimensional analysis up to an overall numerical factor. As a consequence, even though the width itself diverges at small spatial momenta, particles remain narrow resonances and the theory is weakly coupled (at least, as far as this particular process goes), provided the dimensionless coupling
$g/\mu^5$ is small. 

It is natural to expect  that the dispersion relation (\ref{w=m/k})  makes decoupling of
heavy particles subtle compared to the standard case. As an illustration
consider a situation when in addition to a light field of mass $\mu$, there is another one
with mass $M \gg \mu$.
Normally, at energies below $M$ heavy particles simply never get produced.
Consequently,
one can write an effective low-energy theory of light fields only, where all the memory about heavy particles present in the full microscopic theory is
absorbed into the values of coupling constants describing local interactions of the light fields.
This is no longer the case with the UV/IR mixing dispersion relation (\ref{w=m/k}). For example, a two body decay
\be
\label{heavy2body}
\ket{\mu,k} \to \ket{\mu,p} \ket{M,q}
\ee
with $|k| \sim \mu$ and  $|p|,|q| \sim k M^2/\mu^2 $ is kinematically allowed no matter how heavy the mass $M$ is.
Consequently, in a strict sense, the full decoupling never takes place and one can always produce arbitrarily heavy particle simply by waiting long enough.

However, if the coupling constants for light and heavy fields are of the same order, there is an approximate notion of decoupling. One indication for this comes from the possibility to perform a Wick rotation, which for instantaneous theories amounts to replacing $x^+\to i x^+$. The resulting ``Euclidean" propagator $\l \omega^2k^2+\mu^4\r^{-1}$ does not have poles just like it happens in ordinary theories.
Similarly to the conventional case for calculating Green's functions involving light fields only at low values of kinematic invariants one can Taylor
expand the heavy propagators in $\omega^2k^2/M^4$, which trades a heavy field for an infinite number of local vertices involving light fields only---the standard ``integrating out" procedure.

For low energy processes involving on-shell heavy particles, such as the two-body decay (\ref{heavy2body}),
the suppression comes from the difference in the wave function normalization  in
(\ref{field}) between heavy and light fields\footnote{This suppression is not unrelated to the Euclidean argument above. At the end of the day,
the propagator is built as a convolution of two wave functions.}. Replacing an external line of a light field by a heavy one in a given Feynmann diagram brings in a suppression both because of the heavier mass in the denominator of (\ref{field}), and because the corresponding spatial momentum is higher as the example (\ref{heavy2body}) illustrates.

To further illustrate this reasoning   let us consider a super-renormalizable (and finite) theory of three scalar fields $\phi_1$, $\phi_2$ and $\phi_M$ with masses $\mu$, $\mu$, and  $M$ satisfying the hierarchy $\mu\ll M$. As an interaction potential we chose
\[
V_{int}={g_{12}}\phi_M^2\l \phi_1+\phi_2\r+g(\phi_1^3+\phi_2^3)\;,
\] 
so that different light fields interact only through the exchange of the heavy field.
Note first that the decay of a light particle into two heavy particles has a width
\[
\Gamma_{lhh}\simeq {g_{12}^2\over \mu^2M^6 k}\;,
\] 
where we suppressed all order one factors. Consequently, in order for this process to be subleading with respect to
the processes involving light fields only one needs the condition
\be
\label{12cond}
g_{12}\ll g {M^3\over\mu^3}%\ll M^2\mu^2
\ee
to be satisfied. 
%The last inequality in (\ref{12cond}) follows from requiring for the light physics itself to be perturbative.

 In particular, as anticipated, processes with heavy particles are indeed suppressed if $g_{12}\sim g$.
In this case, just like in conventional theories, we can integrate out the heavy field, which will result in direct interactions
between different light fields $\phi_1$, $\phi_2$. At the leading order in $g_{12}$ these are mass, kinetic and higher order in momenta mixings of the form
\be
 g^2_{12}M^{-6-4n} \phi_1 (\p\m)^{2n}\phi_2\;,
 \label{operators}
\ee
which are small as a consequence of (\ref{12cond}) and the smallness of $g/\mu^5$. 

To summarize, we see  that in many situations decoupling in instantaneous theories works similarly
to the standard case, with the important difference that with enough time one will always be able to produce arbitrarily heavy particles on-shell.
On the other hand, the condition (\ref{12cond}) is not necessary for the theory to be perturbative (for instance, one can set $g=0$). If it is violated, one ends up being in an interesting situation when the dominant processes involving low-energy light particles are decays into heavy particles.

%-----------------------------------------------------------------------------------------------------------------------------------

\section{Turbulence}
\label{kinetical}
After developing some experience with simplest interacting instantaneous theories, let us come back to the observation at the end of Section~\ref{classical} that thermalization is impossible in these models and describe
quantitatively 
 the turbulent cascade corresponding to the Bose condensation onto an infinite momentum ground state, which was anticipated there.

The first simple thing to try is just to study numerically the evolution of some smooth initial data in the interacting $\phi^3$ theory. The typical result
\begin{figure}[t!] %  figure placement: here, top, bottom, or page
 \begin{center}
 \includegraphics[width=3.4in]{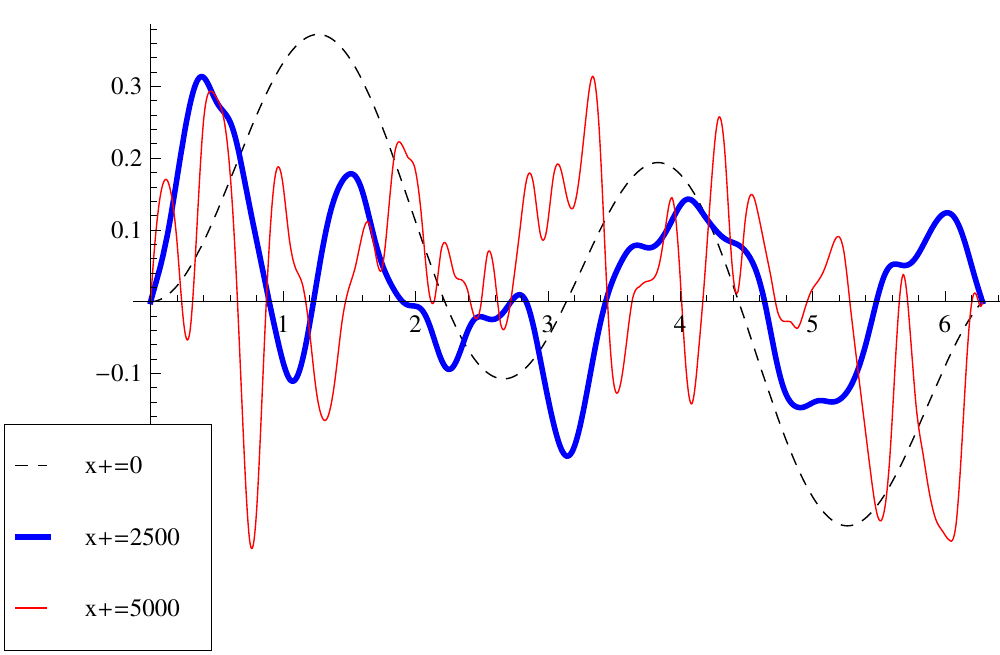}
 \caption{Solution of the classical equation of motion for the $\phi^3$ theory with boundary conditions $\phi \vert_{x_-=0}=\phi \big|_{x_-=2 \pi}=0$ and with $\mu=g=1$.}
 \label{fig:noise}
 \end{center}
\end{figure}
is shown in Fig.~\ref{fig:noise}. As expected, we indeed observe that at late time the  field configuration turns into a short-distance noise.
To describe the statistical properties of this noise quantitatively we make use of the kinetic approach. Namely, we consider  a homogeneous (on average) gas of $\phi$-particles with spatial momenta distribution described by the distribution function $n(k)$. The evolution of this distribution function
is determined by the Boltzmann equation, where in the limit of weak non-linearities the collision integral contains two-body decays, and the inverse process---two-to-one collisions.
 A particle with a momentum $k$ participates
in six processes
%\be
%\begin{eqnarray}
%  { k} \to -k/\varphi,k/\varphi^2  &
%   \frac{g^2}{\mu^8 k}  \left( \varphi+\frac{1}{\varphi} \right)^{-1} n(k)(n(-k/\varphi)+1)(n(k/\varphi^2)+1)  \\
%   \varphi^2k \to k,-\varphi k &
%   \frac{g^2}{\mu^8 k}  \left( \varphi+\frac{1}{\varphi} \right)^{-1} n(\varphi^2k)(n(k)+1)(n(-\varphi k)+1)  \\
 %  \varphi^2k \to k,-\varphi k &
%   \frac{g^2}{\mu^8 k}  \left( \varphi+\frac{1}{\varphi} \right)^{-1} n(\varphi^2k)(n(k)+1)(n(-\varphi k)+1)  \\
 %  \ldots  & \ldots
%\end{eqnarray}
%\ee
\be
   k \leftrightarrow -\varphi k, \varphi^2 k \quad
  %\ket{ k} \leftrightarrow \ket{-\varphi k} \ket{ \varphi^2 k} \quad
    \varphi^{-2} k \leftrightarrow k, -\varphi^{-1} k  \quad
   -\varphi^{-1} k \leftrightarrow k, -\varphi k \quad
\ee
  The $k\to pq$ decay  contributes to the  collision integral as
\be
  -\frac{g^2}{8\mu^8 k}  \left( \varphi+\frac{1}{\varphi} \right)^{-1} n(k)(n(p)+1)(n(q)+1)
\ee
and the inverse processes as
\be
  \frac{g^2}{\mu^8 k}  \left( \varphi+\frac{1}{\varphi} \right)^{-1} (n(k)+1)n(p)n(q)\;.
\ee
The resulting Boltzmann equation reads
\begin{eqnarray}
  \d_{\tau} n(k,\tau) &=& \frac{1}{k} \Big( -n(k,\tau)+n(-\varphi^{-1} k, \tau) +n(\varphi^{-2} k,\tau) -  \nonumber \\
      & &                                                   -n(\varphi^2k,\tau)n(k,\tau)+
                                                    n(\varphi^2k,\tau)n(-\varphi^{-1} k,\tau)
                                                   -2n(-\varphi k,\tau)n(k,\tau)+ \nonumber \\
      & &                                      +  n(\varphi^{-2} k,\tau)n(k,\tau)
                                                        +n(\varphi^{-2} k,\tau)n(-\varphi^{-1} k,\tau)+n(-\varphi^{-1} k,\tau)n(-\varphi k,\tau)\Big)
\label {Boltz}
\end{eqnarray}
%n[t, i] + g*\[Beta]^(i) (-n[t, i] + n[t, i - 1] + n[t, i - 2] -
 %   nl (n[t, i + 2] n[t, i] - n[t, i + 2] n[t, i + 1] + 
    %   2 n[t, i + 1] n[t, i]
   %    - n[t, i - 2] n[t, i] - n[t, i - 2] n[t, i - 1] - 
    %   n[t, i - 1] n[t, i + 1]))
where we absorbed the dependence on the coupling constant into the definition of the time variable 
\[
\tau= \frac{1}{2 \pi}
  \frac{g^2}{8\mu^8}  \left( \varphi+\frac{1}{\varphi} \right)^{-1}x_+\;.
\]

\begin{figure}[t!] %  figure placement: here, top, bottom, or page
 \begin{center}
 \includegraphics[width=3.4in]{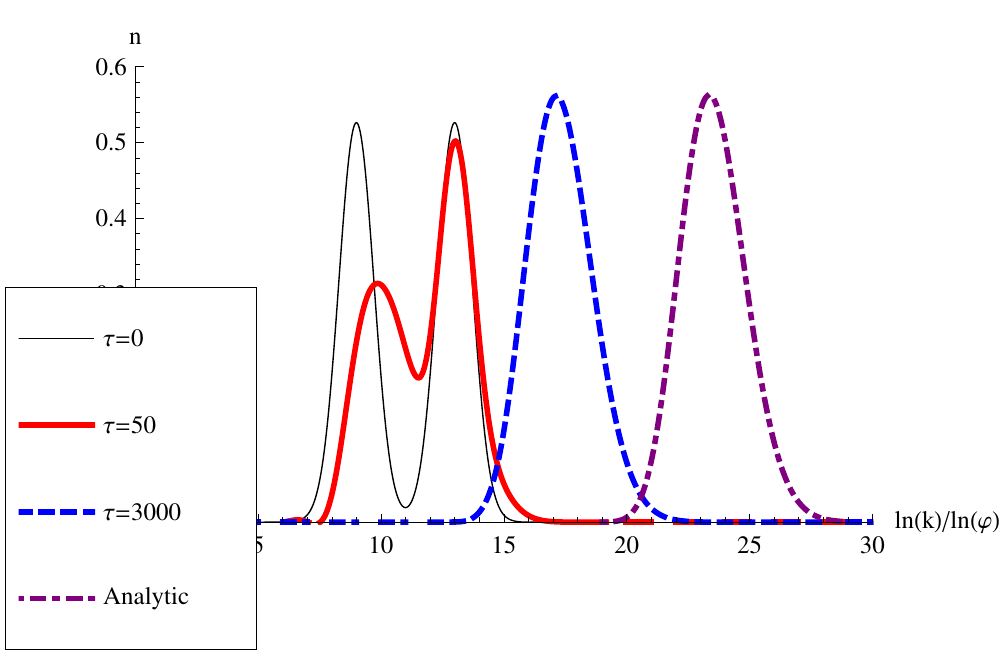}
 \includegraphics[width=3.4in]{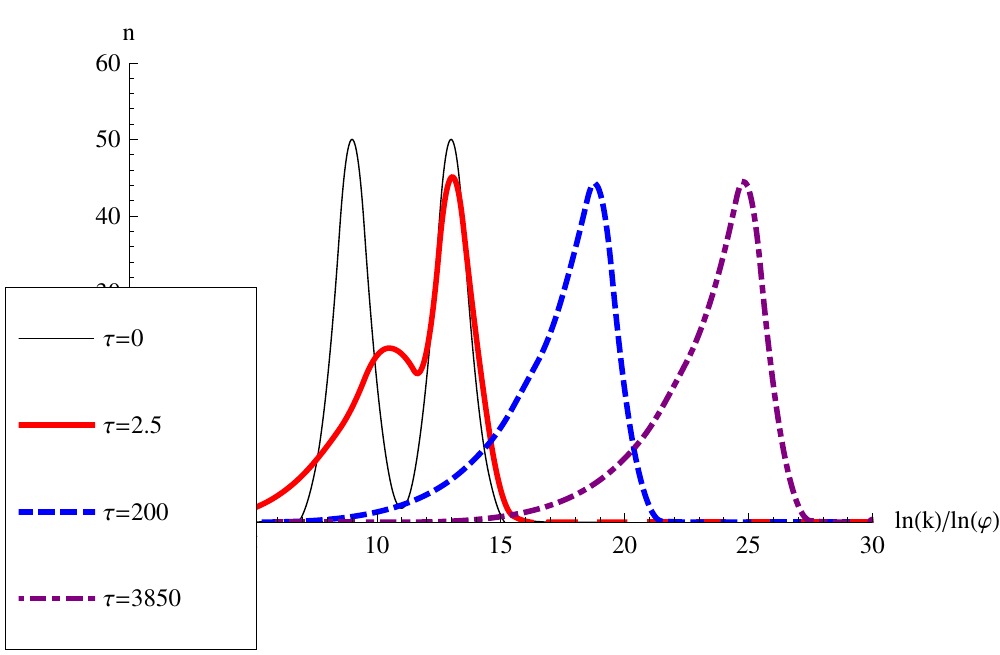}
 \caption{Time evolution of the distribution function as a function of $\log k$ for the  linearized (left) and full (right) Boltzmann equation. In both cases the distribution function converges to the universal shape at late times, which in the linearized case well agrees with the analytic solution.}
 \label{fig:turb}
 \end{center}
\end{figure}

Let us start by considering a low density gas with all occupation numbers being small\footnote{Note, that this is a quantum regime, which cannot be captured by solving the classical field equations.}, $n(k) \ll 1$. 
 In this regime one can drop quadratic contributions to  the collision term in  (\ref{Boltz}) and all the dynamics is determined just by the particle decays, 
\be
   \d_{\tau} n(k,\tau)=
  \frac{1}{k} \left( -n(k,\tau)+n(-\varphi^{-1} k, \tau)+n(\varphi^{-2} k,\tau) \right )\;.
  \label{BoltzLin}
\ee
This equation turns out to be amenable to the analytic study.
For simplicity we restrict to the case $n(k)=n(-k)$.
It is natural to look for a special solution of (\ref{BoltzLin}) of the form 
\be
\label{n*}
n_0(k,\tau)=f\l\log{k\over \tau}\r\;.
\ee
Then the function  $f(x)$ satisfies the following equation
\be
  -e^x f'(x)=-f(x)+f(x-\log\varphi)+f(x-2\log\varphi)\;.
  \label{Boltzf}
\ee 
By performing the Fourier transform we reduce the problem to finding an analytic function decaying along the real axis
and satisfying the following quasi-periodicity condition  
\be
\label{quasip}
  (-iy-1)\hat{f}(y-i) =(1-\varphi^{iy+1})(1+\varphi^{iy-1})\hat{f}(y)\;.
\ee
In the absence of the $\varphi$ dependent prefactors on the r.h.s. of (\ref{quasip})
this would be a defining property of the $\Gamma$-function, $\Gamma(-iy)$. The extra prefactors are straightforward to account for 
by writing the solution in the form
\[
\hat{f}(y)=\Gamma(-i y)P_+(y)P_-(y)\;,
\]
where $P_\pm$ are two uniformly converging infinite products, 
\be
  P_+(y)=\prod_{m=0}^{\infty}\l 1-\varphi^{iy-m}\r, \quad P_-(y)=\prod_{m=2}^{\infty}\l 1+\varphi^{iy-m}\r%
\ee 
which can also be expressed through the  q-Pochhammer symbols as $P_\pm=(\pm\varphi^{\pm1+iy},\varphi)_\infty/(1\mp\varphi^{\pm1+iy})$.
Note, that all the poles of the Gamma function get cancelled by the zeroes of $P_+$.
Gamma function also guarantees the convergense of the inverse Fourier transform at the infinity.
This all implies that the function
\be
  f(x)=\int e^{-iyx}\Gamma(-iy)P_+(y)P_-(y) dy
\ee
satisfies the equation (\ref{Boltzf}) and decays at the infinity faster than 
$e^{-a|x|}$ for any $a$.
Clearly this function is real, however in order to correspond to a physical 
 solution of (\ref{Boltzf}) it also has to be positive.
Fortunately, as the plot in Fig.~\ref{fig:turb}a) shows, 
$f(x)$ is indeed positive and, consequently, provides a solution to the linearized Boltzmann equation via (\ref{n*}).
Moreover, the same plot demonstrates that this solution is actually a universal attractor, describing the shape of the distribution function
as a function of $\log k$ at late stages of the cascade independently of the initial shape (of course, there is a freedom in choosing an overall normalization, which is fixed by the total energy of the gas).

To understand the origin of this attractor behavior note that 
by applying constant time shifts to $n_0$ we can construct the whole family of solutions of the form $n_u(k,\tau) = f(\log k-\log (\tau+u))$.
All these solutions have the same bell-like shape as a function of $\log k$, and differ only by an overall shift.
Given an arbitrary initial profile $n_i(k)$ at $\tau=0$ we can then decompose it in a linear combination of  $n_u(k,0)$, with $u>0$. The corresponding solution of the Boltzmann equation takes then the following form
\be
\label{intsol}
n(k,\tau)=\int_0^\infty du f(\log k-\log\l \tau+u\r) c(u)
\ee
where $c(u)$ are the coefficients in the linear decomposition of $n_i(k)$. For the initial configuration to have a finite total density of particles
the function $c(u)$ should decay at the infinity faster than $1/u$. Then the late time asymptotics of the solution is 
\[
n(k,\tau\to\infty)=f\l\log{k\over\tau}\r\int_0^\infty du \, c(u)
\]
which is exactly the attractor behavior exhibited in Fig.~\ref{fig:turb}a).
  
The same argument cannot be applied for the full non-linear Boltzmann equation (\ref{Boltz}). However, as illustrated in Fig.~\ref{fig:turb}b),
numerical results indicate that the attractor behavior persists at the non-linear level as well. Independently of the initial shape at late time the distribution function is described by a universal scaling solution $f_{nl}(\log{k\over \tau}, E)$, where $E$ is a total energy. Heuristically, the origin of this attractor behavior can be understood as a consequence of the rate of interactions getting lower as the momenta of the particles grow. As a consequence, for any initial distribution, ``fast" low momenta particles always have enough time to catch up with ``slow" high momenta modes and to establish a stable shape, which keeps drifting into UV afterwards.

%------------------------------------------------------------------------------------------------------------------------
\section{Discussion and future directions}
\label{discussion}
To conclude, we hope to have convinced the reader that instantaneous theories 
present a tractable example of apparently local Lorentz invariant quantum field theories exhibiting interesting non-local features. 
Our analysis of their dynamics is far from being exhaustive, we focussed on the simplest case of two-body decays. It would be interesting to study more  general processes, in particular one may expect peculiar IR singularities in scattering processes following from the absence of the gap in the spectra of instantaneous fields.

As we said in the Introduction, the main motivation to insist on Lorentz invariance is that it provides a natural way to couple these models to gravity.
It will be interesting to study whether consistent instantaneous quantum gravity models can be constructed and what is the fate of black holes in these theories.

Note also, that a two dimensional version of the ghost condensate model \cite{ArkaniHamed:2003uy} becomes power counting renormalizable if promoted to an instantaneous theory
\be
{\cal L}_{gc}=(\Box\phi)^2+P\l (\d\phi)^2\r\;.
\ee
It will be interesting to see whether this setup provides a useful lesson for understanding the dynamics of the ghost condensate and more general phases of massive gravity \cite{Dubovsky:2004sg}.
Dynamics of these theories may be somewhat more subtle than in examples studied in this paper, because the shift symmetry acting on $\phi$
prevents one from adding a mass term $\mu^4\phi^2$.

As a totally different direction, note that one may view the theories studied here as a specific case of non-relativistic theories with an anisotropic scaling symmetry 
\[
\tau \to \lambda \tau,\;\sigma \to \lambda^{z}\sigma.
\]
Theories of this kind with $z<0$ are likely to share many of the features of $z=-1$ case studied here. It is an interesting question whether anisotropic scaling symmetries with $z<0$ may arise in condensed matter system. In this regard it is worth noting that the dispersion relation
(\ref{w=m/k}) describes short-wavelength nondivergent Rossby waves in the Earth atmosphere in the absence of mean currents \cite{Rossby}.

Finally, the physics of instantaneous models relies crucially on  special properties of a two-dimensional Lorentz group. It is not clear whether the lessons learnt here may be applied in higher dimensions. A promising proposal in this direction \cite{superlum} may be to study the dynamics of a string, with the instantaneous causal structure on its world-sheet, propagating in a higher dimensional space-time. We hope to report a progress on this and other related topics in a near future.
\section*{Acknowledgements}
We thank Giga Gabadadze, Andrei Gruzinov, Sergey Paston and Arkady Vainshtein for useful discussions.
This material is based upon work supported in part by the National Science Foundation under Grants No. 1066293 and NSF PHY05-51164 and the hospitality of the Aspen Center for Physics
and of the Kavli Institute for Theoretical Physics at UC  Santa Barbara.

\end{document}